# Nickel antidot arrays on anodic alumina substrates

Z. L. Xiao, Catherine Y. Han, U. Welp, H. H. Wang, V. K. Vlasko-Vlasov, W. K. Kwok,
D. J. Miller, J. M. Hiller, R. E. Cook, G. A. Willing,  and G. W. Crabtree

Materials Science Division, Argonne National Laboratory, Argonne, IL 60439

Large area nickel antidot arrays with density up to $10^{10}$ /cm$^2$ have been fabricated by depositing nickel onto anodic aluminum oxide membranes that contain lattices of nanopores.  Electron microscopy images show a high degree of order of the antidot arrays.  Various sizes and shapes of the antidots were observed with increasing thickness of the deposited nickel. New features appear in the antidot arrays in both magnetization and transport measurements when the external magnetic field is parallel to the current direction, including an enhancement and a nonmonotonous field dependence of the magnetoresistance, larger values of the coercive field and remanence moment, and smaller saturation field.

Recently, a new generation of ultrahigh density magnetic storage media has attracted much attention. Promising candidates include arrays of magnetic nanowires [1], self-assembled magnetic nanoparticles [2], magnetic dots [3,4] and antidots [5-12]. In the latter case antidots are believed to have advantages over dots [5]. First of all, there is no superparamagnetic lower limit to the bit size because there is no isolated volume; the stability of the written bits increases with increasing storage density in the antidot array rather than stays unchanged as in the dot array. Although the antidot array can be formed by self-assembly during the film deposition [6] or annealing the films after fabrication [7], a common and more controllable method is to pattern the films using electron-beam lithography [5, 8-11] or focused ion beam milling [12]. Submicrometer antidots have been demonstrated using these advanced techniques. In this Letter we present a way to fabricate ultrahigh density magnetic antidots with sizes down to 20 nm by depositing magnetic material on substrates with highly ordered arrays of nanopores. The magnetic antidot arrays were characterized using field emission scanning electron microscopy (FESEM), transmission electron microscopy (TEM), magnetization and magnetoresistance measurements.

The substrates used are anodic aluminum oxide (AAO) membranes formed through anodizing aluminum foils in an acidic solution. Arrays of pores of various diameters and spacings can be fabricated in AAO membranes by choosing appropriate anodization conditions [13]. Compared to other types of membranes containing arrays of nanopores, e.g. nuclear track-etched mica [14] and polycarbonates [15], and micelles [16], AAO membranes have unique properties characterized by the excellent uniformity in diameter and spacing of the pores. Our AAO membranes were fabricated by a two-step anodization procedure as described previously [13]. The starting materials were aluminum foils with dimensions of 20x30x0.5 mm$^3$. In order to achieve a surface with a roughness in the range of a few nanometers, the foils were first electropolished in a mixed solution of perchloric acid and ethanol (1:8) for 10 minutes at a current density of 200-500 mA/cm$^2$. The first anodization was carried out in a 0.3M oxalic acid solution at 0°C by applying a constant voltage of 40 V for 24 hours. The resulting alumina layer was then removed by immersing the specimen in a mixture of phosphoric acid (6wt%) and chromic acid (1.8wt%) at room temperature. This procedure leaves a highly ordered array of dimples on the aluminum surface that initializes the pore array formation in the following anodization. The second anodization was carried out under the same condition as the first one. After the second anodization AAO membranes with highly ordered arrays of nanopores were obtained by removing the unreacted aluminum in a saturated HgCl$_2$ solution. The diameter of the as-grown pores is about 40 nm and can be enlarged by immersing AAO membranes in phosphoric acid (5wt%). Because the size of the magnetic antidots is related to the stability of the written bits, we fabricated antidots of various sizes with the same density by depositing magnetic material of various thickness on membranes with pore size of about 70 nm.

Figure 1 (a) shows a FESEM (Hitachi S-4700-II) image of a typical pore array in an AAO membrane anodized at 40 V followed by 30 minutes pore widening in phosphoric acid. In order to avoid charging effect during imaging, a thin nickel layer of about 5 nm was deposited on the surface. The uniformity of the pore diameter and the high degree of order of the array can be seen. The spacing between the centers of the pores is about 100 nm. This corresponds to a density of pores of $10^{10}$/cm$^2$ or a storage density of 10 Gbits/cm$^2$.

The magnetic antidots were prepared by depositing nickel onto AAO membranes at room temperature by DC magnetron sputtering. The base vacuum was better than $10^{-4}$ Pa. High purity argon was used as working gas at a pressure of 2.5-3.0 Pa. The deposition rate was about 10 nm/min. Simultaneously with the Ni/AAO samples reference Ni-films were deposited on glass substrates. Inspection by FESEM the shape and size of the antidots changed with increasing thickness of the deposited nickel. When the nickel layer was thin (5 nm) the antidots retained the same shape and size as those of the pores in the AAO membrane. With increasing thickness of nickel, the size of the holes was reduced and the shape of the holes became hexagonal at a nickel thickness of about 40-50 nm [see Fig.1(b)]. Further increase of the nickel thickness to 100 nm resulted in an irregular shape of holes [see Fig.1(c)] of about 20 nm diameter. The holes finally closed and a continuous film formed at a thickness of about 200 nm. The inset of Fig. 1(c) shows the cross section of the 100-nm film obtained after cleaving the sample and imaging at an angle of 45 degree to surface of the membrane. It shows that the antidots indeed have open bottom and the size of the antidots is the largest at the bottom and gradually decreases toward the top surface. These results demonstrate a highly controlled method varying the size and shape of the antidots which have important consequencces on the stability of the written bits [5]. TEM images reveal that the Ni-films are polycrystalline with an average grain size of about 10-15 nm. Dark-field TEM showed that there is no preferential grain alignment.

The magneto-transport and magnetization data for the 100 nm thick films are summarized in Fig. 2. The magnetic field was applied in the film plane either parallel or perpendicular to the current direction, yielding the longitudinal and transverse magneto resistance, respectively. The continuous film is characterized by an almost reversible magnetization (Fig. 2c) which is for both orientation essentially the same and approaches saturation near 0.15 T, and by a monotonous negative magnetoresistance, MR = $R(H)/R(0)-1$ (Fig. 2a). The magneto-transport data can be accounted for in the frame work of conventional anisotropic magnetoresistance (AMR) [17] in which an anisotropic spin-orbit coupling induces maximum resistance when the magnetic moments are aligned parallel to the current direction and a minimum resistance for perpendicular alignment. The AMR can be described by $R = R_t + (R_l - R_t) \cos^2(\ )$. Here, $R_t$ and $R_l$ are the transverse and longitudinal resistance obtained when extrapolating the magnetoresistance from saturation back to H = 0 (as indicated in Fig. 2), and    is the angle between magnetization and current. The AMR ratio $(R_l - R_t)/R(H=0)$    0.2 % is reduced as compared to values on bulk samples [18]. This may have the following reasons. In constrained geometries such as thin films or thin wires the magnetoresistance is suppressed due to scattering at the sample surfaces [18]. In addition, due to the polycrystalline nature of the samples field independent contributions to the resistivity arising from grain-boundary scattering reduce the AMR ratio. A weak, almost linear and orientation independent magnetoresistance occurs in fields well above saturation. Similar behavior is frequently observed and is attributed to the suppression of spin-scattering [19].

The magnetization curve of the antidot array exhibits a loop with enhanced values of coercive field and remnant magnetic moment (see Fig. 2c). We attribute these changes to the interplay of shape anisotropy and inhomogeneous magnetization rotation caused by the nano-scale patterning of the magnetic film. For bulk Ni the domain wall width is $\delta = (A/K)^{1/2} \approx 125$ nm with the exchange constant A=8x10$^{-7}$ erg/cm [20] and the anisotropy constant $K$= 5x10$^4$ erg/cm$^3$ [21]. In samples composed of nano-scale non-textured grains the effective anisotropy is strongly reduced [22] as compared to the bulk value and $\delta$ is enhanced. Thus, in the patterned Ni-films studied here the effective domain wall width is substantially larger than the bridges between the holes and, hence the nucleation and propagation of magnetic domain walls is not expected. The magnetization process then occurs through the inhomogeneous rotation of the magnetic moments. One could also expect that the magnetoelastic anisotropy which is introduced by the film/substrate lattice mismatch and is usually essential in Ni films due to large magnetostriction coefficients [23] should be less in the antidot array due to the large free surface. A lower saturation field H$_s$ for the antidot array compared to the continuous film in Fig.2(c) supports such a suggestion. In addition, the magnetostatic energy, 2$\pi$M$_s^2$, associated with the large internal surfaces induces the preferential alignment of the magnetic moments parallel to the hole circumference giving rise to the enhanced low-field magnetization hysteresis (Fig.2c).

The magneto-transport data of the patterned 100 nm thick Ni-film are shown in Fig. 2b. The hysteretic behavior, particularly in the longitudinal resistivity component, is clearly seen. Whereas the transverse magnetoresistance shows similar over-all behavior and magnitude as seen in the continuous film, unusual non-monotonic behavior is observed in the longitudinal magnetoresistance which we attribute to the inhomogeneous rotation of magnetic moments with respect to the applied field and current directions. Since the remnant moment is small in low magnetic fields, the magnetic moments point in all directions. However, the moments as well as the current trajectories, conform to the hole array. Therefore, locally, current and magnetization are largely parallel or antiparallel resulting in an enhanced longitudinal AMR. With increasing field the magnetic moments inhomogeneously and irreversibly rotate towards the field direction whereas the current flow pattern does not change. Transverse components of the magnetization (with respect to the current direction) arise causing the observed decrease in the resistivity. With further increasing fields, these transverse components decrease again and the magnetoresistance increases. Correspondingly, in the transverse resistance (the average current direction is perpendicular to the field) these transverse components are essentially parallel to the current and contribute to a higher resistance. Superimposed on the steep field dependence a shoulder near 0.06 T (marked by the vertical dotted lines in Fig. 2b) signals this effect which is then suppressed with increasing field, in good agreement with the longitudinal data.

In summary, we demonstrated an attractive way to achieve ultrahigh density recording media by depositing magnetic layers onto substrates with arrays of ultrahigh density nanopores. Arrays of nickel antidots on porous anodic aluminum oxide membranes show high degree of order and adjustability of the size and shape of the antidots. New features in the field dependence of the magnetization and magnetoresistance indicate opportunities of interesting new physics in nanoscale antidot arrays.

This work was supported by the US Department of Energy (DOE), BES-Materials Science, Contract No. W-31-109-ENG-38. The FESEM imaging was performed in the Electron Microscopy Center of Argonne National Laboratory.

**Figure captions**

Fig.1. Field emission scanning electron microscopy (FESEM) images of the nickel antidot arrays on anodic aluminum oxide (AAO) substrates. The thickness of the nickel layers is 5 nm (a), 50 nm (b) and 100 nm (c), respectively. The inset of (c) is the side view of the nickel antidot array of 100 nm thick.

Fig.2. Field dependence of the transverse and longitudinal magnetoresistances of the 100 nm thick Ni-on-glass reference sample (a) and of the 100 nm thick Ni-on-AAO sample (b). The magnetoresistance MR is defined as: MR=R(H)/R(0)-1. Comparison of the magnetization of both samples (c). The inset in (c) shows the low-field magnetization on expanded scales. The measurements were carried out at a temperature of 280 K.

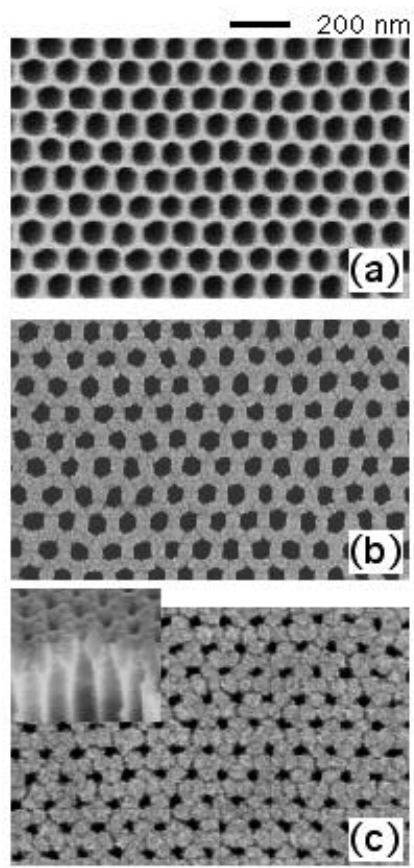

200 nm

(a)

(b)

(c)

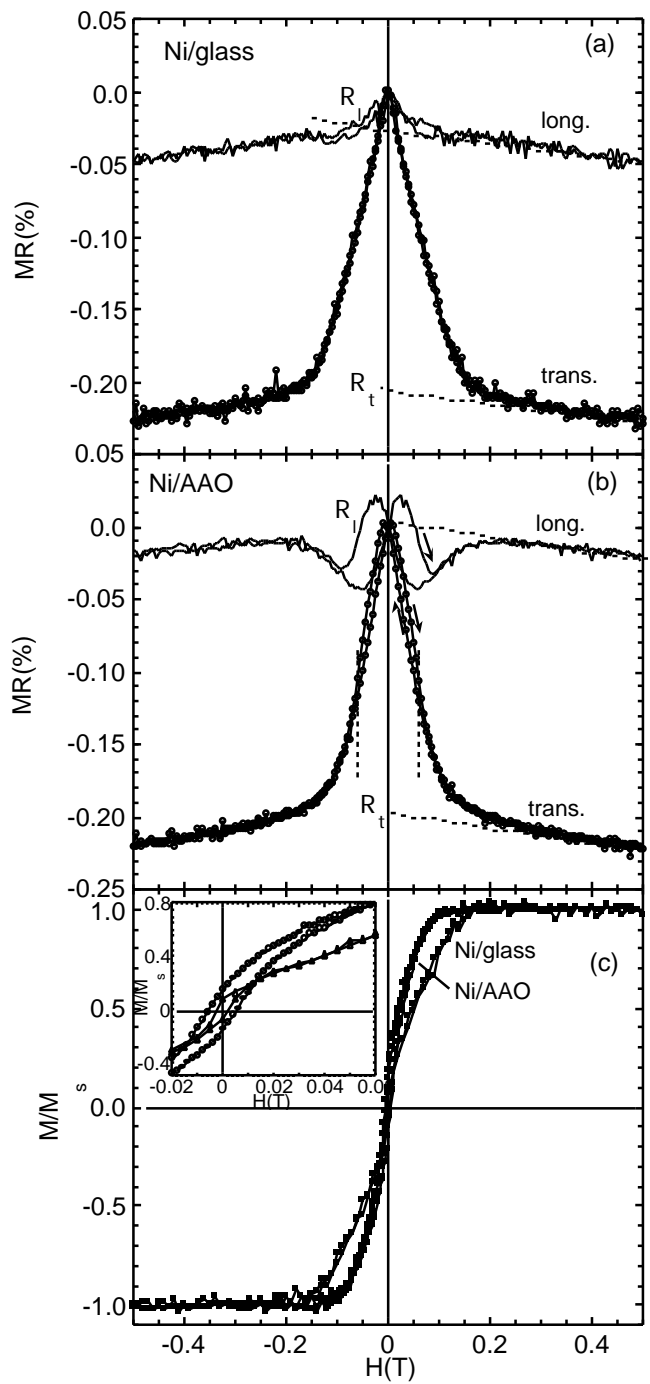

Fig. 2
Z. L. Xiao et al.